\documentclass[referee,sn-nature]{sn-jnl}

\usepackage{graphicx}%
\usepackage{multirow}%
\usepackage{braket}
\usepackage{amsmath,amssymb,amsfonts}%
\usepackage{amsthm}%
\usepackage{mathrsfs}%
\usepackage[title]{appendix}%
\usepackage{xcolor}%
\usepackage{textcomp}%
\usepackage{manyfoot}%
\usepackage{booktabs}%
\usepackage{algorithm}%
\usepackage{algorithmicx}%
\usepackage{algpseudocode}%
\usepackage{listings}%
\usepackage{float}
\usepackage{hyperref}
\usepackage{cleveref}
\usepackage{hyperref}
\usepackage{lineno}

\usepackage{chngcntr}
\usepackage{ulem}
\usepackage[separate-uncertainty=true]{siunitx}

\theoremstyle{thmstyleone}%
\theoremstyle{thmstyletwo}%

\theoremstyle{thmstylethree}%

\begin{document}

\title[Article Title]{Nonperturbative Nonlinear Transport in a Floquet-Weyl Semimetal}


\author[1,2]{\fnm{Matthew W.} \sur{Day }}
\author[1,2]{\fnm{Kateryna} \sur{Kusyak}}
\author[1,2]{\fnm{Felix} \sur{Sturm}}
\author[3]{\fnm{Juan I.} \sur{Aranzadi}}
\author[2]{\fnm{Hope M.} \sur{Bretscher}}
\author[2]{\fnm{Michael} \sur{Fechner}}
\author[2]{\fnm{Toru} \sur{Matsuyama}}
\author[2]{\fnm{Marios H.} \sur{Michael}}
\author[1,2]{\fnm{Benedikt F.} \sur{Schulte}}
\author[2]{\fnm{Xinyu} \sur{Li}}
\author[2]{\fnm{Jesse} \sur{Hagelstein}}
\author[2]{\fnm{Dorothee} \sur{Herrmann}}
\author[2]{\fnm{Gunda} \sur{Kipp}}
\author[2]{\fnm{Alex M.} \sur{Potts}}
\author[4]{\fnm{Jonathan M.} \sur{DeStefano}}
\author[4]{\fnm{Chaowei} \sur{Hu}}
\author[1]{\fnm{Yunfei} \sur{Huang}}
\author[5]{\fnm{Takashi} \sur{Taniguchi}}
\author[6]{\fnm{Kenji} \sur{Watanabe}}
\author[1]{\fnm{Guido} \sur{Meier}}
\author[7]{\fnm{Dongbin} \sur{Shin}}
\author[2,8,9]{\fnm{Angel} \sur{Rubio}}
\author[4]{\fnm{Jiun-Haw} \sur{Chu}}
\author[8]{\fnm{Dante M.} \sur{Kennes}}
\author[3]{\fnm{Michael A.} \sur{Sentef}}
\author*[1,2]{\fnm{James W.} \sur{McIver}}
 \affil*{Corresponding author}

\affil[1]{\orgdiv{Department of Physics}, \orgname{Columbia University}, \orgaddress{\city{New York, NY}, \country{USA}}}
\affil[2]{\orgname{Max Planck Institute for the Structure and Dynamics of Matter},\orgaddress{\city{ Hamburg}, \country{Germany}}}
\affil[3]{\orgdiv{Institute for Theoretical Physics and Bremen Center for Computational Materials Science}, \orgname{University of Bremen}, \orgaddress{\city{Bremen},\country{ Germany}}}
\affil[4]{\orgdiv{Department of Physics},\orgname{University of Washington},\orgaddress{\city{Seattle}, \country{USA}}}

\affil[5]{\orgdiv{Research Center for Materials Nanoarchitectonics}, \orgname{National Institute for Materials Science}, \orgaddress{\city{Tsukuba},\country{ Japan}}}
\affil[6]{\orgdiv{Research Center for Electronic and Optical Materials}, \orgname{National Institute for Materials Science}, \orgaddress{\city{Tsukuba},\country{ Japan}}}
\affil[7]{\orgdiv{Department of Physics and Photon Science}, \orgname{Gwangju Institute of Science and Technology}, \orgaddress{\city{Gwangju},\country{ South Korea}}}
\affil[8]{\orgdiv{Center for Computational Quantum Physics}, \orgname{Simons Foundation Flatiron Institute}, \orgaddress{\city{New York, NY},\country{ USA}}}
\affil[9]{\orgdiv{CNano-BioSpectroscopy Group, Departamento de Fisica de Materiales}, \orgname{Universidad del País Vasco}, \orgaddress{\city{San Sebastián},\country{ Spain}}}

\affil[10]{\orgdiv{Institut für Theorie der Statistischen Physik}, \orgname{RWTH Aachen University and JARA-Fundamentals of Future Information Technology}, \orgaddress{\city{Aachen},\country{ Germany}}\vspace{2cm}}

\abstract{Periodic laser driving, known as Floquet engineering, is a powerful tool to manipulate the properties of quantum materials. Using circularly polarized light, artificial magnetic fields, called Berry curvature, can be created in the photon-dressed Floquet-Bloch states that form. This mechanism, when applied to 3D Dirac and Weyl systems, is predicted to lead to photon-dressed  movement of Weyl nodes which should be detectable in the transport sector. The transport response of such a topological light-matter hybrid, however, remains experimentally unknown. Here, we report on the transport properties of the type-II Weyl semimetal T$\mathrm{_d}$-MoTe$_\mathrm{2}$ illuminated by a femtosecond pulse of circularly polarized light. Using an ultrafast optoelectronic device architecture, we observed injection currents and a helicity-dependent anomalous Hall effect whose scaling with laser field strongly deviate from the perturbative laws of nonlinear optics. We show using Floquet theory that this discovery corresponds to the formation of a magnetic Floquet-Weyl semimetal state. Numerical \textit{ab initio} simulations support this interpretation, indicating that the light-induced motion of the Weyl nodes contributes substantially to the measured transport signals. This work demonstrates the ability to generate large effective magnetic fields ($>$ 30T) with light, which can be used to manipulate the magnetic and topological properties of a range of quantum materials.}


\maketitle

\section{Introduction}\label{sec1}

Tailored ultrafast laser fields can selectively manipulate the macroscopic properties of quantum materials away from equilibrium \cite{Basov2017,AnkitAndrea2020,delatorre2021,Bao2022,Borsch2023}. Leveraging a range of light-matter coupling mechanisms \cite{delatorre2021}, experiments have demonstrated the optical control of magnetism \cite{Ankit2023,Afanasiev2021,Kimel2019,Raising2010}, ferroelectricity \cite{LiNelson2019,TaoLi2018,RubioMarcos2018,Basov2017}, topology \cite{JamesBenedikt2020,Yoshikawa2022}, charge density waves \cite{Xing2024,Zong2018,Yoshikawa2021,Rettig2021,Vaskivskyi2015}, structural phases \cite{Forst2011,Meredith2022,AnkitAndrea2020,delatorre2021}, electronic motion \cite{Galan2020,Huber2023,Borsch2023}, and superconducting-like metastable phases \cite{Mitrano2016,Budden2021}. 

At the frontier of optical control paradigms is Floquet engineering, whereby time-periodic laser driving creates photon-dressed Floquet-Bloch states \cite{Oka2009,Kitagawa2010,Lindner2011,Wang2014,Gedik2016,Sentef2017,Huber2023}. The properties of these states and their transport responses can be manipulated \textit{in situ} by adjusting parameters of the light field. For example, when using circularly polarized light, which coherently breaks time-reversal symmetry (TRS), artificial magnetic fields known as Berry curvature can be created or manipulated \cite{Oka2009,Oka2018,Rudner2020}. Many proposals have leveraged this idea to predict a range of dressed topological phases \cite{Oka2009, Lindner2011,Oka2016}.

Experiments in solids \cite{Gedik2015,Gedik2016, JamesBenedikt2020,Matsunaga2023,Huber2023,Zhou2023,Merboldt2024,Choi2024}, metamaterials \cite{Rechtsman2013}, and cold atoms \cite{Gregor2014,Monica2024} have realized some of these predictions in two-dimensional systems, in which the light-induced Berry curvature can trigger a change in overall topological invariant of the system \cite{Oka2009,Lindner2011}. In three-dimensional systems, more crystalline symmetries can be broken \cite{Zhou2021,delatorre2021,Orth2022}, leading to a wider variety of possible light-induced topological phases \cite{Lindner2011,Shimano2024,Sentef2017,Rudner2020,RodriguesVega2019,Oka2016}, aspects of which have been investigated in optical experiments \cite{Matsunaga2023,Shimano2022,Shimano2023}. Central to these proposals is the concept of the creation and manipulation of photon-dressed Weyl nodes \cite{Wang2014,Sentef2017,Oka2018}, which act as sources and sinks of Berry curvature in momentum space. Light-induced changes in Weyl node position redistributes Berry curvature at the Fermi surface \cite{Chan2016,Sentef2017}, which can be directly sensed in the transport sector \cite{vonKlitzing2020}. However, due to the technical complexities inherent to capturing transport in out-of-equilibrium systems, the transport response of a Floquet-Weyl semimetal remains elusive.  

\begin{figure}
    		\centering
      \renewcommand{\thefigure}{\arabic{figure}}
         \newcommand*{\figuretitle}[1]{%
    {\centering%
    \textbf{#1}%
    \par\medskip}%
}
      
      \vspace*{8mm}
       \makebox[\textwidth][c]{\includegraphics{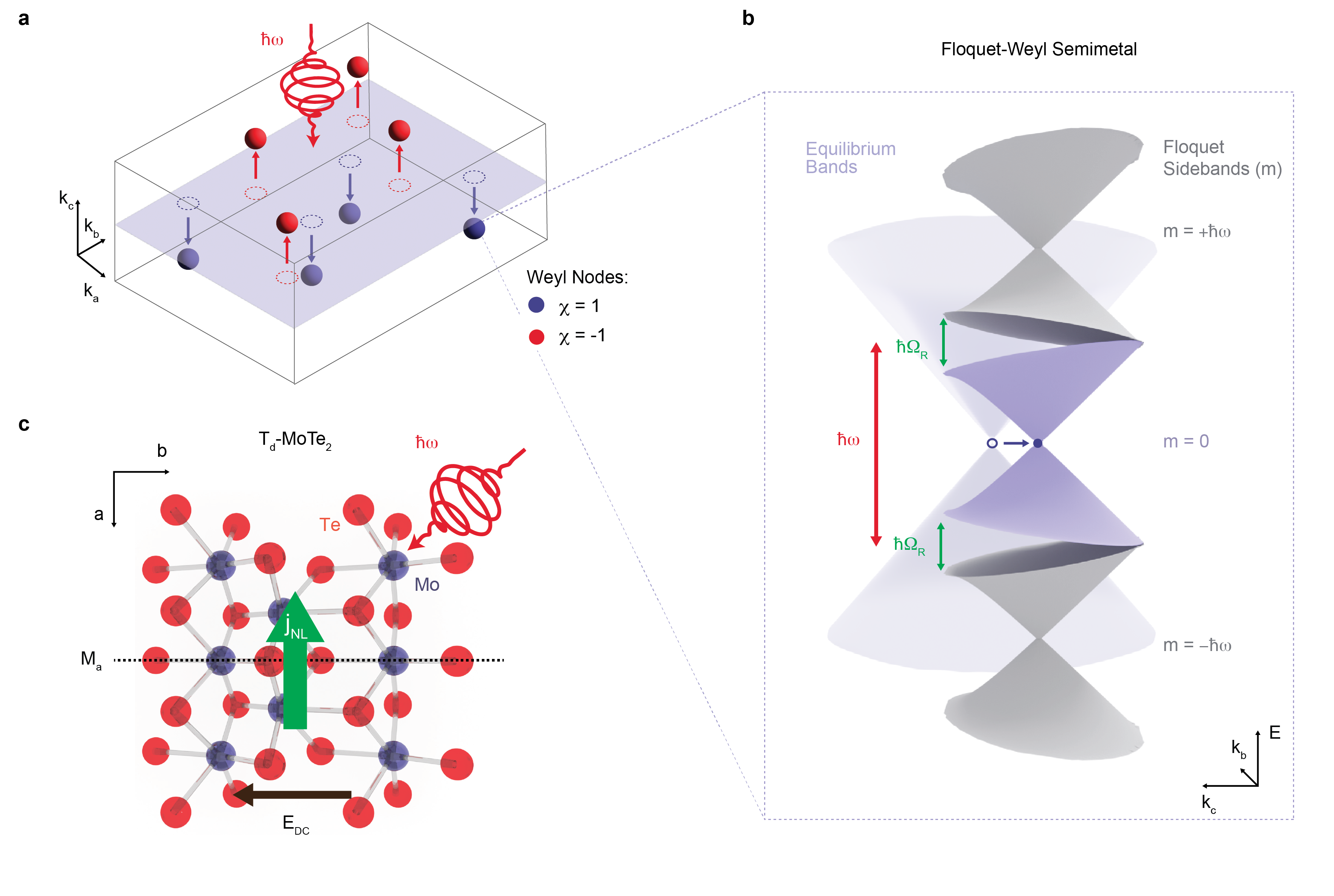}}
    		
      \vspace*{-4mm}
    		\caption{\textbf{Qualitative features of a Floquet-Weyl semimetal a,b: } Depiction of the photon-dressed state (Floquet-Weyl semimetal), formed when illuminating the sample with intense mid-infrared light: the Weyl nodes move in opposite direction in $k_z$, and hybridisation gaps open where the photon-dressed bands intersect. \textbf{c} The surface crystal structure of T$_\mathrm{d}$-MoTe$_\mathrm{2}$ with the green arrow indicating the expected direction of the nonlinear responses under illumination with circularly polarized mid-infrared light. The direction of the transverse bias applied during light-induced Hall measurements is depicted, and is co-directional with the sample mirror plane, M$_a$.} 
		    \label{fig:fig1}
    \end{figure}

Here, we report on the transport properties of the type-II Weyl semimetal T$_d$-MoTe$_2$ driven by a femtosecond pulse of circularly polarised light (Fig. \ref{fig:fig1} a-c). Using an ultrafast optoelectronic device architecture \cite{Auston1975,JamesBenedikt2020}, we observed injection currents and an anomalous Hall effect whose scaling with driving field cannot be interpreted using the perturbative laws of nonlinear optics. We show, using Floquet theory, that the nonperturbative scaling is a natural consequence of Floquet-Bloch sideband formation and hybridisation. \textit{Ab initio} calculations indicate that intense out-of-plane driving with circularly polarized light pushes the two inequivalent Weyl nodes of T$_d$-MoTe$_2$ in opposite directions, with movements of up to $>$5\% of the Brillouin zone at the highest fluences depending on their chirality. Additionally, gaps form at the photon resonances where Floquet-Bloch sidebands hybridise, having gap sizes calculated to be up to $\sim$20\% of the photon energy which indicates the coherent light-matter interaction is in the ultrastrong coupling regime. These photon-dressing effects lead the Floquet-Bloch states to break TRS, creating a photo-induced phase transition to a low-symmetry magnetic Floquet-Weyl state, posessing an effective magnetic field of up to 30 T. This interpretation is supported by the observation of a light-induced anomalous Hall effect that occurs only during the peak of the photocurrent signal \cite{JamesBenedikt2020}.

\section{Ultrafast transport}
\begin{figure}
    		\centering
      \renewcommand{\thefigure}{\arabic{figure}}
         \newcommand*{\figuretitle}[1]{%
    {\centering%
    \textbf{#1}%
    \par\medskip}%
}
      
      \vspace*{8mm}
       \makebox[\textwidth][c]{\includegraphics{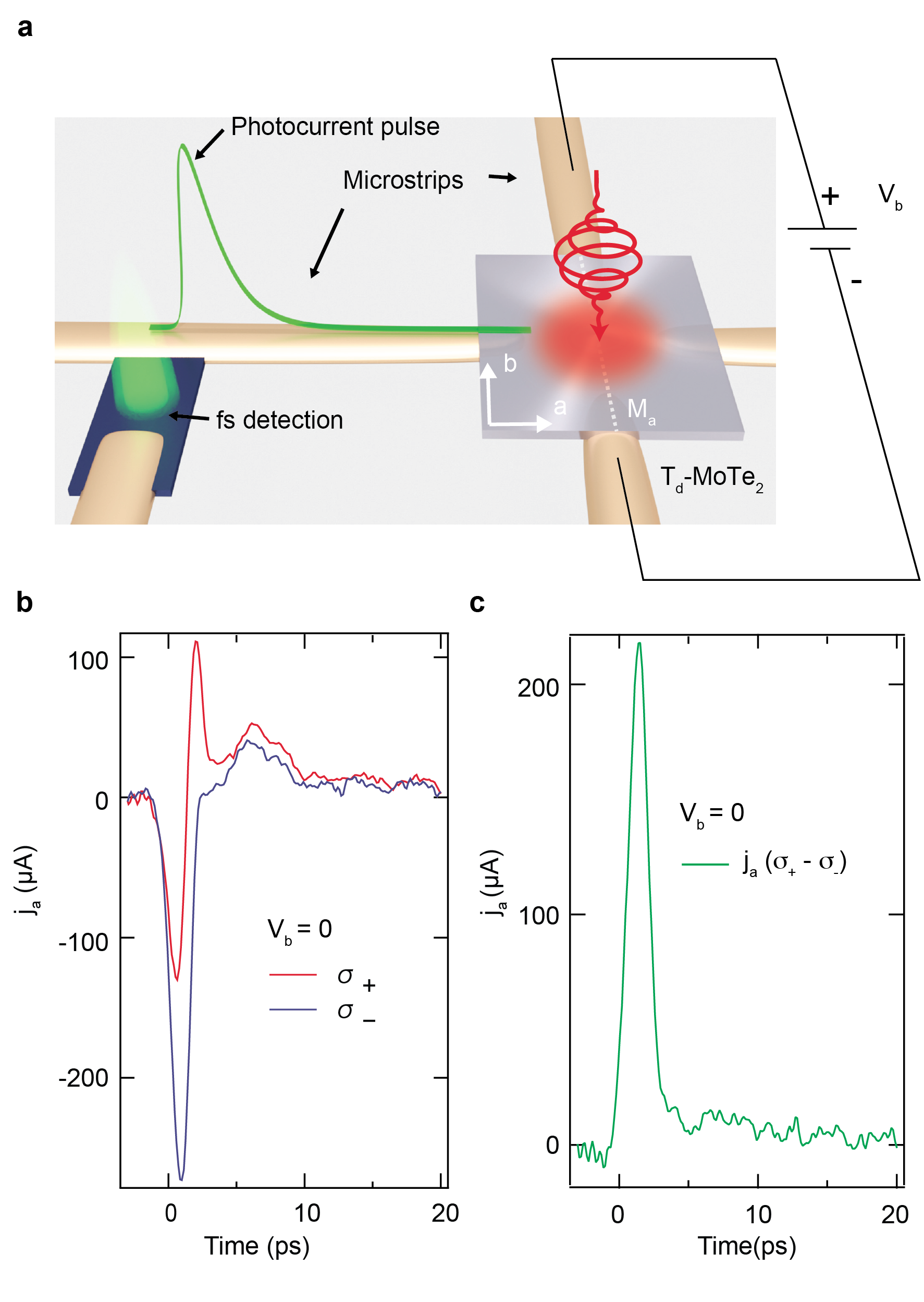}}
    		
      \vspace*{-4mm}
    		\caption{\textbf{Ultrafast transport experiment and time-resolved photocurrents:} \textbf{a} Ultrafast photocurrent circuitry which captures the transient response of a T$_\mathrm{d}$-MoTe$\mathrm{2}$ flake under circularly polarized excitation. \textbf{b} Time-resolved photocurrent stroboscopically measured by adjusting the time delay between excitation and detection pulses using right- and left-handed circularly polarized light, in the absence of transverse DC bias. \textbf{c} The total circularly dichroic signal, taken by subtracting the transient photocurrent measured with left-handed circular light from that measured with right-handed circular light. A monopolar in-plane transient photocurrent is measured only along the \textit{a}-axis \cite{supplement}.}
		    \label{fig:fig2}
    \end{figure}

To study the transport dynamics of Floquet-Bloch states, a technique must be used which can isolate photocurrents generated during the laser pulse, i.e. when the photon-dressed states have formed and are hybridised with the equilibrium bands. We achieved this by ohmically contacting a bulk flake (15~-~30~nm thick) of T$_\mathrm{d}$-MoTe$_\mathrm{2}$, to an ultrafast circuitry architecture based on transmission lines and photoconductive switches triggered by femtosecond laser pulses (Fig. \ref{fig:fig2} \textbf{a}). The crystal axes of the flake were aligned to the transmission lines (see Methods). The flake was driven with ultrafast pulses of mid-infrared ($\lambda \sim$ 10~µm) circularly polarized light with peak fields up to 1.5~MV/cm. We detected time-resolved photocurrent traces at cryogenic temperatures ($\sim 15$ K) with right-handed circularly polarized light and subtracted those taken with left-handed circularly polarized light. This serves as a direct measure of the circularly dichroic, TRS-breaking photocurrent (Fig. \ref{fig:fig2} \textbf{b} and \textbf{c}) which emanates from the sample perpendicular to the mirror plane along the \textit{b}-axis, denoted M$\mathrm{_a}$. By temporally resolving the photocurrent transients, intrinsic photocurrent dynamics and excitation mechanisms during the laser drive can be disentangled from slower photocurrent generating effects \cite{Song2023,supplement}. 

Along the sample \textit{a}-axis, we observed a net, bias-free charge transfer under optical illumination indicative of a circularly dichroic injection current (Fig. \ref{fig:fig2} \textbf{c}, and \cite{supplement}). To determine the microscopic origin of this photocurrent, we first consider the symmetry of thin T$_\mathrm{d}$-MoTe$_\mathrm{2}$, which belongs to the Pmm$_{11}$ space group \cite{Loh2020} (corresponding to the C$_\mathrm{1v}$ point group), and thus has a proper mirror plane along the \textit{b}-axis (Fig. \ref{fig:fig1} \textbf{a}, \ref{fig:fig2} \textbf{a}). Under these symmetry constraints, the only symmetry-allowed net in-plane circularly dichroic injection current driven with out-of-plane light is perpendicular to this mirror plane \cite{Sun2021} (i.e. along \textit{a}-axis), in agreement with our data \cite{supplement}. Using \textit{ab initio} computational methods, we identify the bands contributing to this bulk photogalvanic effect by calculating the optical selection rules for circularly polarized light of appropriate wavelength impinging on the sample \cite{supplement}. We attribute the microscopic origin of this injection current to the combination of optical selection rules around the Weyl nodes with the asymmetric band velocities in this inversion-symmetry-broken system. Such a photocurrent are emblematic of a circular photogalvanic effect (CPGE), whereby direct interband excitations take electrons from the valence to conduction bands and asymmetries in band velocities combine with these selection rules to result in net directional photocurrents \cite{Soifer2019,Song2023}. 

\section{Nonperturbative injection currents}
\begin{figure}[h]
    		\centering
      \renewcommand{\thefigure}{\arabic{figure}}
         \newcommand*{\figuretitle}[1]{%
    {\centering
    \textbf{#1}%
    \par\medskip}
}
      \vspace*{8mm}
       \makebox[\textwidth][c]{\includegraphics{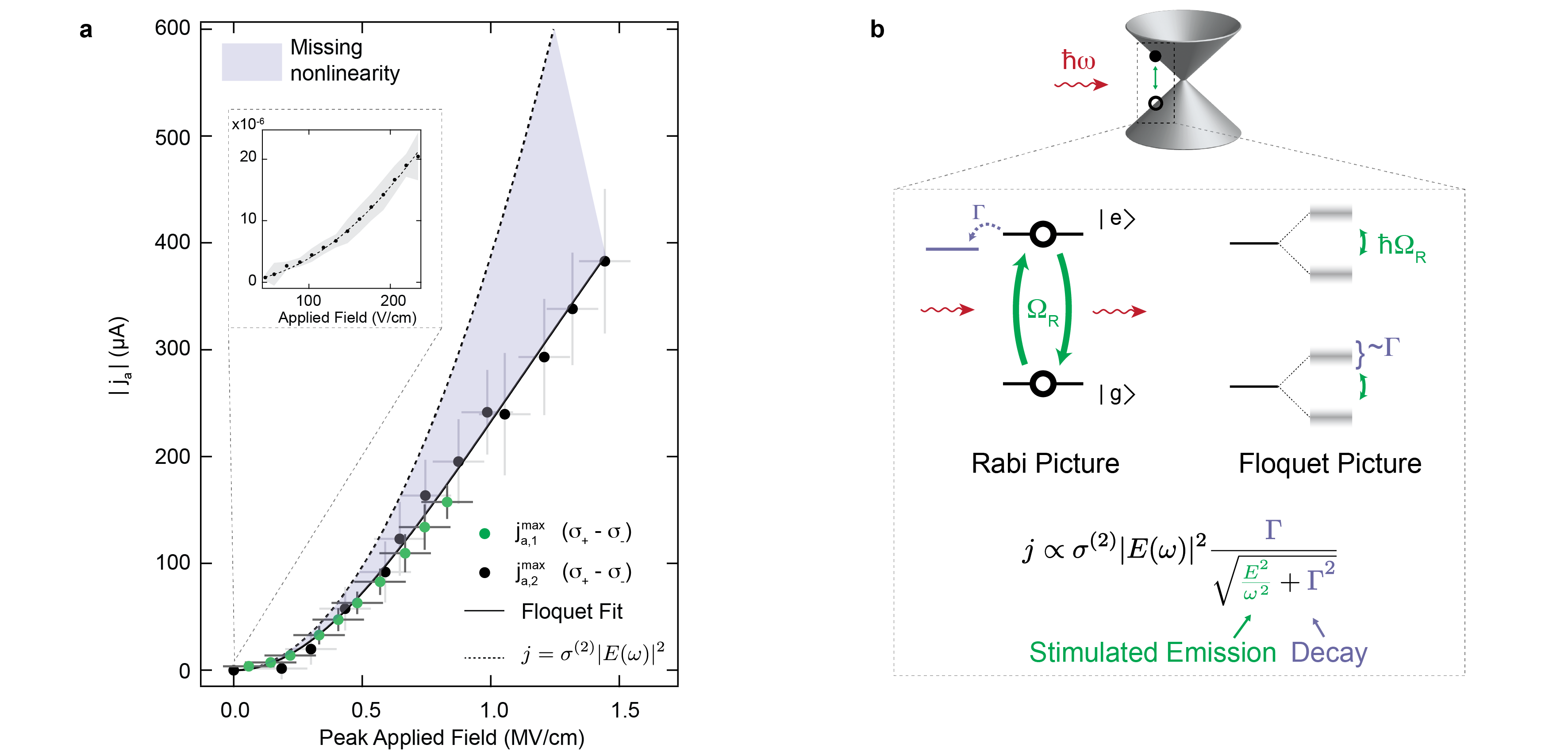}}
      \vspace*{-4mm}
    		\caption{\textbf{Nonperturbative nonlinear transport in the Floquet regime: a} The peak of the circularly dichroic photocurrent, measured in two devices, scales Nonperturbatively with excitation field strength. The dotted line is a fit to low-fluence data, extended into this Nonperturbative regime. The data diverge strongly from this fit around a peak field of 0.5 MV/cm. \textbf{b} The qualitative picture for the missing nonlinearity is that during excitation, the system begins to synchronize with the optical field, resulting in stimulated emission which reduces the overall photocurrent yield. In the time-domain, this can be thought of as Rabi oscillations occurring for each vertical optical transition where the electrons emit photons on the down-flop of the Rabi cycle. In the frequency domain (Floquet picture) this corresponds to a k-dependant optical Stark effect. }
		    \label{fig:fig3}
    \end{figure}

We measured the helicity-dependent injection current at a variety of field strengths to capture the crossover between the perturbative and nonperturbative regimes. We began by using a low-field continuous wave laser centered at 10 µm. We acquired the time-averaged circularly dichroic photocurrent along the \textit{a}-axis as a function of laser field \cite{supplement}. The amplitude of the measured CPGE signal in the low-field regime depends quadratically on excitation field, as can be seen in Fig. \ref{fig:fig3}. This quadratic scaling is what one would expect from the established understanding of photocurrent generation in semimetalic systems, fit using Eq. ~\ref{eq:1} \cite{Chan2017a,Sun2021,Golub2022,Song2023}. This fit is depicted as the dashed line in Fig. ~\ref{fig:fig3} \textbf{a}. Such an observed scaling at low field strengths accords well with the simple expectation from nonlinear response theory: the number of photons absorbed scales quadratically with the electric field, and therefore so should the photocurrent. Explicitly, the CPGE response of a material can be calculated using

\begin{equation}
    \frac{d j_a}{d t} = \beta_{ac}(\textbf{E}(\omega) \times \textbf{E}^*(\omega))_c
    \label{eq:1}
\end{equation}
where $\beta_{ac}$ is the ac-th component of the CPGE tensor (directly proportional to the ac-th component of the interband Berry curvature dipole \cite{SuYang2018,Liuyan2021}), and $\textbf{E}(\omega)$ is the in-plane diving field applied by our pulse illuminating at normal incidence.

However, as the driving field was increased by four and a half orders of magnitude, the peak of the time resolved CPGE begins to strongly diverge from perturbative scaling, as seen in Fig.~\ref{fig:fig3} \textbf{a}. At the highest peak fields ($\sim$ 1.5 MV/cm), we find, remarkably, that the CPGE scales linearly in the electric field, in violation of the perturbative expectation of nonlinear response theory. To our knowledge, this kind of anomalous peak photocurrent field-scaling from an intrinsic injection current transient has never been observed. 

This observation can be interpreted by treating the response within the Floquet picture. Following the theoretical framework developed in Refs.~\cite{Sodeman2021} and \cite{Nagaosa2016}, rectified photovoltaic currents in the nonperturbative regime can be described by: 

\begin{equation}
    j = a(\Gamma, \textbf{E}) \frac{|\textbf{E}|^2 }{\sqrt{\frac{|\textbf{E}|^2}{\omega^2} + \frac{\Gamma^2}{4} b(\Gamma,\textbf{E})^2}}.
    \label{eq:2}
\end{equation}

Here, $a(\Gamma,\textbf{E})$ and $b(\Gamma,\textbf{E})$ are constants that depend on material details and experimental geometry (See \cite{supplement}), while $\Gamma$ parameterises the timescale associated with non-radiative decay processes, such as scattering from impurities or phonons. The factor of (E/$\omega$)$^2$ parametrises stimulated emission into the radiation field \cite{Sodeman2021,Nagaosa2016}. In the language of nonlinear response theory, this corresponds to the photon-dressed bands opening another relaxation channel for the excited-state electrons, similar to excited-state emission in atoms. As the peak field of the laser increases, the excited state population begins to synchronize with the light field, undergoing Rabi oscillations in the time domain. In the frequency domain, the Rabi oscillations manifest as the hybrdizaiton gaps at the photon resonances as depicted in Fig. ~\ref{fig:fig3} \cite{Sodeman2021,Matsunaga2023}. 

The gaps depicted in Figs. \ref{fig:fig1},\ref{fig:fig3}, and \ref{fig:fig4} can be understood as the \textit{k}-dependent complement to the AC Stark effect where a light field can hybridise with an atomic energy level \cite{Gedik2015,Bakos1977}, creating photon-dressed replicas at integer multiples of the photon energy. This hybridisation between the photon-dressed and original eigenstates forms an avoided crossing, opening a gap parameterised by the Rabi frequency \cite{Bakos1977}. Moreover, the coherence between the illuminating field and oscillations of electronic population in the material provides an additional scattering channel for the excited electrons. These electrons do not contribute to the overall photocurrent because they maintain their coherence with the driving field. The additional ‘reservoir’ provided to the excited electrons by the coherent light-matter oscillations (Rabi oscillations) limits the nonlinear scaling of the rectified response of the sample to be exactly linear in the driving field at high field strengths \cite{Sodeman2021}. 

We fit the data to a unit-restored version of the expression in Eq. \ref{eq:2} where we account for the reduction in the expected light-matter coupling due to the field reduction inside a semimetal \cite{supplement}. This fit has only two free parameters: an overall scaling factor, and the scattering rate of free carriers in the sample. In the model in Ref.~ \cite{Sodeman2021}, such a scattering rate is analytically derived from a tight-binding Hamiltonian describing carriers coupling to a fermionic bath, but qualitatively captures incoherent removal of electrons from the excited state independent of the excitation field. From the fit, we extract the incoherent carrier scattering rate, $1/\Gamma$ = 90$\pm$30 fs, in agreement with that calculated from the width of the bands as measured by angle-resolved photoemission spectroscopy \cite{Grioni2017,BiswasSoren2021}, but an order of magnitude shorter than the scattering rate derived from magneto-transport measurements \cite{supplement}. This is the factor up to which CPGE is quantised in Weyl semimetals, and is difficult to otherwise extract when operating in the perturbative regime \cite{Moore2017}. Because this scattering rate is roughly three times longer than a single optical cycle, coherence between the optical field and sample is established, as corroborated by time-resolved, angle-resolved photo-emission spectra taken in similar materials at similar field strengths \cite{Gedik2016,Huber2023}.

\section{Light-induced anomalous Hall effect}

\begin{figure}[H]
    		\centering
      \renewcommand{\thefigure}{\arabic{figure}}
         \newcommand*{\figuretitle}[1]{%
    {\centering
    \textbf{#1}%
    \par\medskip}%
}
      
      \vspace*{8mm}
       \makebox[\textwidth][c]{\includegraphics{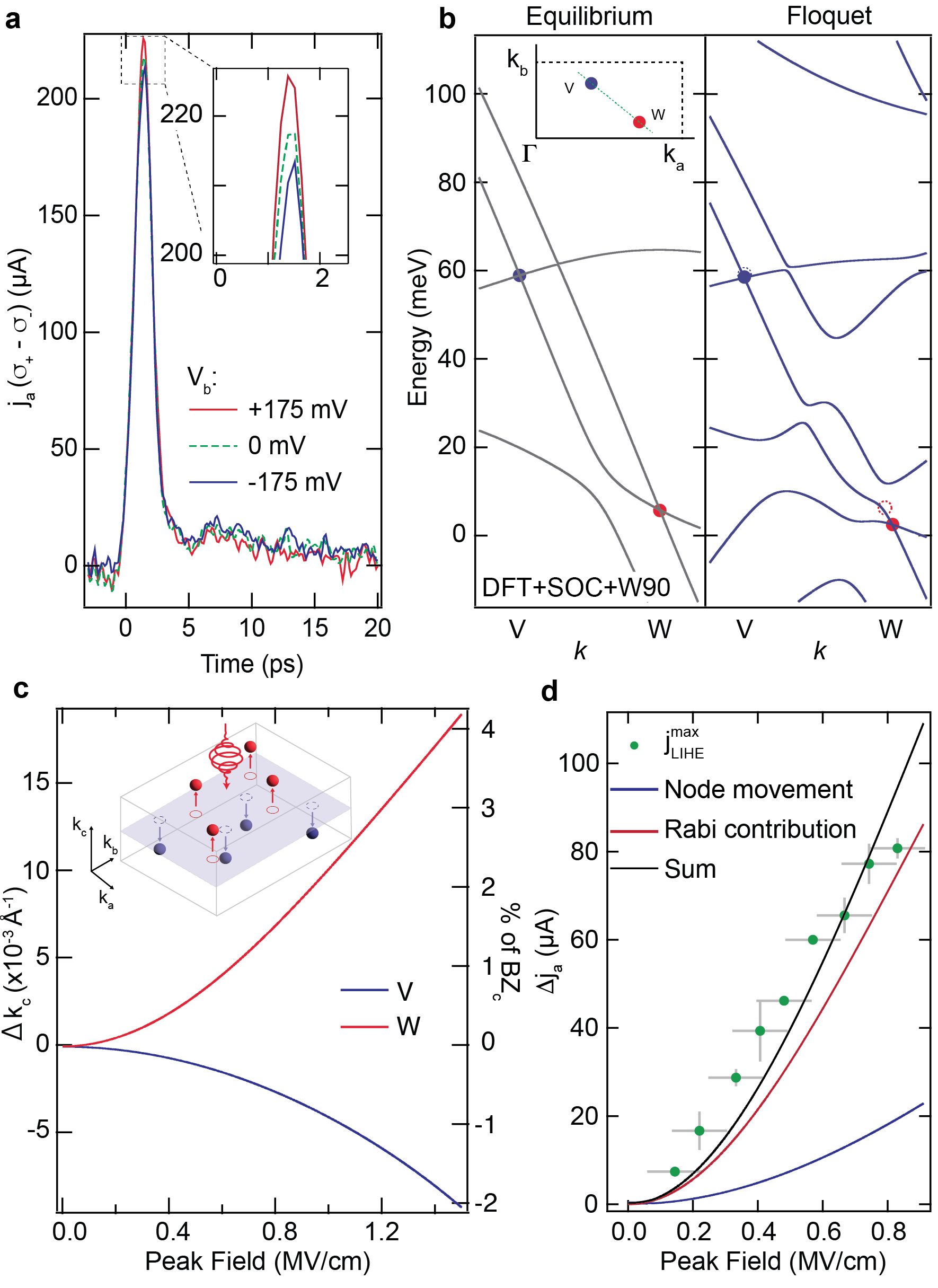}}
    		
      \vspace*{-4mm}
    		\caption{\textbf{Non-equilibrium light-induced anomalous Hall effect: a} Transient photocurrent traces measured under different transverse bias scenarios overlap except at the very peak of the signal, corresponding to a transient light-induced Hall effect (inset). \textbf{b} Bandstructure (left) and photon-dressed bandstructure under optical driving with a field strength of 1.5MV/cm and photon energy of 135 meV (right). \textbf{c} The Weyl nodes move away from the $k_z = 0$ plane in polarisation-dependent directions, corresponding to the formation of a transient Floquet Weyl semimetal. \textbf{d} Measured LIHE peak as a function of excitation field. To characterize the different contributions to this signal, the contribution to the LIHE from the motion of the Weyl nodes is calculated, and then subtracted from the experimental data. Then, the node motion-free LIHE is fit with a scale-only fit corresponding to the solid line in Fig.~\ref{fig:fig3} \textbf{a} to parameterise the Rabi contribution to the LIHE, coming from carriers at the photon resonances interacting with the bias field.}
		    \label{fig:fig4}
    \end{figure}

Because circularly polarized light is used to create the photon-dressed state, this TRS-broken light-matter hybrid should possess a transient, light-induced anomalous Hall effect (LIHE) \cite{JamesBenedikt2020}, which we observe and is detailed in Fig.~\ref{fig:fig4}. We conduct this measurement by biasing the sample along the \textit{b}-axis and monitoring the circularly dichroic current along the \textit{a}-axis, depicted in \ref{fig:fig2} \textbf{a}. In the orthogonal configuration, we observe no LIHE due to the presence of the mirror plane (i.e., no net dichroic current flows along the \textit{b}-axis). A LIHE is evident in Fig.~\ref{fig:fig4} \textbf{a} as a signal on the peak of the photocurrent trace (Fig.~\ref{fig:fig4} \textbf{a} inset). To understand this signal, we performed \textit{ab initio} density-functional theory calculations which we then Wannierized to extract the single-particle tight-binding Hamiltonian \cite{pizzi2020}. This was then used to calculate the realistic Floquet bandstructure of T$_\mathrm{d}$-MoTe$_\mathrm{2}$ \cite{supplement}. Our calculations show that this photon-dressed state will possess both asymmetric hybridisation gaps and shifted Weyl nodes, qualitatively depicted in Fig.~\ref{fig:fig1} \textbf{b} and \textbf{c}, quantitatively shown in Fig.~\ref{fig:fig4} \textbf{b} and \textbf{c} \cite{Chan2016,Oka2016,Zhou2021}. Breaking both time-reversal and inversion symmetry simultaneously leads to a distinct phase, corresponding to a light-induced transient magnetic Weyl semimetal \cite{Orth2022}.

Two leading-order contributions to the LIHE are considered here: the motion of the Weyl nodes under illumination with circularly polarized light, and the coupling between the bias field and the anisotropic excited state population at the photon resonances \cite{Chan2016,Sato2019,Mathey2020,Shimano2022, Shimano2023,Matsunaga2023}. We therefore measured the LIHE as a function of peak excitation field to estimate the relative contribution of each of these effects. We find that the LIHE scales  linearly in the excitation field at the highest field strengths. To interpret this data, we return to our calculations of the Weyl node motion, shown in Fig. \ref{fig:fig4} \textbf{c}, which also begin to scale linearly at the highest fields. As shown in Ref. \cite{Chan2016}, the opposing motion of the nodes $\delta q$ in $k$-space gives a contribution to the LIHE given by

\begin{equation}
    \delta \sigma_{ij} = \frac{e^{2}}{2 \pi h} \epsilon_{ijk} \sum_{I} \chi^{(I)} \delta q_k^{(I)},
    \label{eq:sigmaxy}
\end{equation}
where the subscript $I$ in the summation runs over the 8 Weyl nodes in the first Brillouin zone, and $\chi$ is the corresponding chirality. The motion of the nodes, shown in Fig. \ref{fig:fig4}, contributes roughly one-fifth of the observed LIHE, as estimated using calculations detailed in \cite{supplement}.

The remaining LIHE signal is likely due to the hybridisation gaps at the photon resonance, which has been shown to contribute prominently to LIHE responses in Dirac semimetals \cite{Chan2016,Sato2019,Mathey2020,Shimano2022, Shimano2023,Matsunaga2023}. We parameterise the photon-resonance contribution to the data in Fig.~\ref{fig:fig4} \textbf{c} by making the ansatz that the photon resonance LIHE contribution scales with the number of excited carriers extracted from the device at peak field, which is in turn proportional to the CPGE at high fields. This contribution to the LIHE is then quantified by subtracting the LIHE generated by the motion of the Weyl nodes from the data and rescaling the fit in Fig.~\ref{fig:fig3} \textbf{a} by a constant chosen to best fit the data after subtracting the Hall effect contribution from the Weyl node motion at the Fermi surface. We denote the photon-resonance contribution to the LIHE as the `Rabi contribution' as depicted in Fig. \ref{fig:fig4} \textbf{d}. At the highest field strengths, we calculate the total effective magnetic field \cite{Merlin2023}, $B^{eff}_{tot} \approx 30$ T, which would yield the same transverse bias, whereas the effective magnetic field from only the motion of the equilibrium Weyl nodes alone is $7$ T \cite{supplement}.

\section{Conclusion and Outlook}

We measured the nonperturbative nonlinear transport properties of a Floquet-Weyl state created by illuminating T$_\mathrm{d}$-MoTe$_\mathrm{2}$ with strong pulses of circularly polarized mid-infrared light. We observed a markedly nonperturbative scaling of the circularly dichroic injection current response that violates the perturbative laws of nonlinear transport, which we interpreted within the Floquet framework as a natural consequence of formation of Floquet-Bloch states. This light-induced state breaks time-reversal symmetry as evidenced by the presence of a light-induced anomalous Hall effect. \textit{Ab initio} calculations indicate that this type of time-reversal broken Floquet state should have the properties of a magnetic Weyl semimetal. 

Our results open the doors to several future directions. The ability to generate such large effective magnetic fields via the creation of Floquet-Bloch states could be harnessed to manipulate magnetic and topological phenomena in a wide range of quantum materials. Additionally, when combined with static or dynamic magnetic fields, or symmetry engineered light which breaks different spatial symmetries, the light-matter hybridisation may result in effective gauge fields for which there is no equilibrium counterpart \cite{Oka2016,Shimano2022}. This type of non-equilibrium driving with tailored light \cite{Orth2022}, when coupled to ultrafast circuitry, could provide experimental access to new non-equilibrium phases in the transport sector. Studying the magneto-transport properties of light-matter hybrids may even lead to a better understanding of Weyl fermionics and chiral anomaly physics \cite{Oka2016}. These experimental and theoretical tools developed in this work can be adapted to a wide array of systems to predict and control their coherent behavior with light.

\backmatter

\bmhead{Acknowledgements}
\noindent This work was primarily supported by the U.S. Department of Energy, Office of Science, Basic Energy Sciences, under Early Career Award DE-SC0024334. The growth and characterization of MoTe$_2$ single crystals was supported by the Center on Programmable Quantum Materials, an Energy Frontier Research Center funded by the U.S. Department of Energy (DOE), Office of Science, Basic Energy Sciences (BES), under award DE-SC0019443. The Max Planck—New York City Center on Non-Equilibrium Quantum Phenomena provided facility support and fellowship support for B.S. and K.K. M.W. D. and M.H.M. acknowledge support from the Alex von Humboldt postdoctoral fellowship. H.M.B. was supported by the H2020 Marie Sklodowska-Curie actions of the European Union (grant agreement no. 799408). J.M.D was supported by the National Science Foundation Graduate Research Fellowship Program under Grant No. DGE-2140004. Any opinions, findings, and conclusions or recommendations expressed in this material are those of the authors and do not necessarily reflect the views of the National Science Foundation. K.W. and T.T. acknowledge support from acknowledge support from the JSPS KAKENHI (Grant Numbers 20H00354 and 23H02052) and World Premier International Research Center Initiative (WPI), MEXT, Japan. A.R. acknowledges support from the Cluster of Excellence ‘CUI: Advanced Imaging of Matter’- EXC 2056 - project ID 390715994 and Grupos Consolidados (IT1453-22). The Flatiron Institute is a division of the Simons Foundation. D.M.K acknowledges funding by the Deutsche Forschungsgemeinschaft (DFG, German Research Foundation) within the Priority Program SPP 2244 “2DMP” - 443274199 and under Germany’s Excellence Strategy - Cluster of Excellence Matter and Light for Quantum Computing (ML4Q) EXC 2004/1 - 390534769. D.S was supported by the National Research Foundation of Korea (NRF) grant funded by the Korean government (MSIT) (RS-2023-00253716 and No. RS-2024-00333664). M.A.S. was funded by the European Union (ERC, CAVMAT, project no. 101124492).

\bibliography{sn-bibliography}

\begin{thebibliography}{10}
\expandafter\ifx\csname url\endcsname\relax
  \def\url#1{\burl{#1}}\fi
\expandafter\ifx\csname urlprefix\endcsname\relax\def\urlprefix{URL }\fi
\providecommand{\bibinfo}[2]{#2}
\providecommand{\eprint}[2][]{\url{#2}}
\providecommand{\doi}[1]{\url{https://doi.org/#1}}
\bibcommenthead

\bibitem{Basov2017}
\bibinfo{author}{Basov, D.~N.}, \bibinfo{author}{Averitt, R.~D.} \& \bibinfo{author}{Hsieh, D.}
\newblock \bibinfo{title}{{Towards properties on demand in quantum materials}}.
\newblock \emph{\bibinfo{journal}{Nature Materials}} \textbf{\bibinfo{volume}{16}}, \bibinfo{pages}{1077--1088} (\bibinfo{year}{2017}).

\bibitem{AnkitAndrea2020}
\bibinfo{author}{Disa, A.~S.} \emph{et~al.}
\newblock \bibinfo{title}{{Polarizing an antiferromagnet by optical engineering of the crystal field}}.
\newblock \emph{\bibinfo{journal}{Nature Physics}} \textbf{\bibinfo{volume}{16}}, \bibinfo{pages}{937--941} (\bibinfo{year}{2020}).

\bibitem{delatorre2021}
\bibinfo{author}{{de la Torre}, A.} \emph{et~al.}
\newblock \bibinfo{title}{{Colloquium: Nonthermal pathways to ultrafast control in quantum materials}}.
\newblock \emph{\bibinfo{journal}{Reviews of Modern Physics}} \textbf{\bibinfo{volume}{93}}, \bibinfo{pages}{041002} (\bibinfo{year}{2021}).

\bibitem{Bao2022}
\bibinfo{author}{Bao, C.}, \bibinfo{author}{Tang, P.}, \bibinfo{author}{Sun, D.} \& \bibinfo{author}{Zhou, S.}
\newblock \bibinfo{title}{{Light-induced emergent phenomena in 2D materials and topological materials}}.
\newblock \emph{\bibinfo{journal}{Nature Reviews Physics}} \textbf{\bibinfo{volume}{4}}, \bibinfo{pages}{33--48} (\bibinfo{year}{2022}).

\bibitem{Borsch2023}
\bibinfo{author}{Borsch, M.}, \bibinfo{author}{Meierhofer, M.}, \bibinfo{author}{Huber, R.} \& \bibinfo{author}{Kira, M.}
\newblock \bibinfo{title}{Lightwave electronics in condensed matter}.
\newblock \emph{\bibinfo{journal}{Nature Reviews Materials}} \textbf{\bibinfo{volume}{8}}, \bibinfo{pages}{668--687} (\bibinfo{year}{2023}).
\newblock \urlprefix\url{https://doi.org/10.1038/s41578-023-00592-8}.

\bibitem{Ankit2023}
\bibinfo{author}{Disa, A.~S.} \emph{et~al.}
\newblock \bibinfo{title}{Photo-induced high-temperature ferromagnetism in \protect{YTiO$_3$}}.
\newblock \emph{\bibinfo{journal}{Nature}} \textbf{\bibinfo{volume}{617}}, \bibinfo{pages}{73--78} (\bibinfo{year}{2023}).
\newblock \urlprefix\url{https://doi.org/10.1038/s41586-023-05853-8}.

\bibitem{Afanasiev2021}
\bibinfo{author}{Afanasiev, D.} \emph{et~al.}
\newblock \bibinfo{title}{{Ultrafast control of magnetic interactions via light-driven phonons}}.
\newblock \emph{\bibinfo{journal}{Nature Materials}} \textbf{\bibinfo{volume}{20}}, \bibinfo{pages}{607--611} (\bibinfo{year}{2021}).

\bibitem{Kimel2019}
\bibinfo{author}{Kimel, A.~V.} \& \bibinfo{author}{Li, M.}
\newblock \bibinfo{title}{Writing magnetic memory with ultrashort light pulses}.
\newblock \emph{\bibinfo{journal}{Nature Reviews Materials}} \textbf{\bibinfo{volume}{4}}, \bibinfo{pages}{189--200} (\bibinfo{year}{2019}).
\newblock \urlprefix\url{https://doi.org/10.1038/s41578-019-0086-3}.

\bibitem{Raising2010}
\bibinfo{author}{Kirilyuk, A.}, \bibinfo{author}{Kimel, A.~V.} \& \bibinfo{author}{Rasing, T.}
\newblock \bibinfo{title}{{Ultrafast optical manipulation of magnetic order}}.
\newblock \emph{\bibinfo{journal}{Reviews of Modern Physics}} \textbf{\bibinfo{volume}{82}}, \bibinfo{pages}{2731--2784} (\bibinfo{year}{2010}).

\bibitem{LiNelson2019}
\bibinfo{author}{Li, X.} \emph{et~al.}
\newblock \bibinfo{title}{{Terahertz field--induced ferroelectricity in quantum paraelectric SrTiO3}}.
\newblock \emph{\bibinfo{journal}{Science}} \textbf{\bibinfo{volume}{364}}, \bibinfo{pages}{1079--1082} (\bibinfo{year}{2019}).

\bibitem{TaoLi2018}
\bibinfo{author}{Li, T.} \emph{et~al.}
\newblock \bibinfo{title}{Optical control of polarization in ferroelectric heterostructures}.
\newblock \emph{\bibinfo{journal}{Nature Communications}} \textbf{\bibinfo{volume}{9}}, \bibinfo{pages}{3344} (\bibinfo{year}{2018}).
\newblock \urlprefix\url{https://doi.org/10.1038/s41467-018-05640-4}.

\bibitem{RubioMarcos2018}
\bibinfo{author}{Rubio-Marcos, F.} \emph{et~al.}
\newblock \bibinfo{title}{Reversible optical control of macroscopic polarization in ferroelectrics}.
\newblock \emph{\bibinfo{journal}{Nature Photonics}} \textbf{\bibinfo{volume}{12}}, \bibinfo{pages}{29--32} (\bibinfo{year}{2018}).
\newblock \urlprefix\url{https://doi.org/10.1038/s41566-017-0068-1}.

\bibitem{JamesBenedikt2020}
\bibinfo{author}{McIver, J.~W.} \emph{et~al.}
\newblock \bibinfo{title}{{Light-induced anomalous Hall effect in graphene}}.
\newblock \emph{\bibinfo{journal}{Nature Physics}} \textbf{\bibinfo{volume}{16}}, \bibinfo{pages}{38--41} (\bibinfo{year}{2020}).

\bibitem{Yoshikawa2022}
\bibinfo{author}{Yoshikawa, N.} \emph{et~al.}
\newblock \bibinfo{title}{Non-volatile chirality switching by all-optical magnetization reversal in ferromagnetic weyl semimetal \protect{Co$_3$Sn$_2$S$_2$}}.
\newblock \emph{\bibinfo{journal}{Communications Physics}} \textbf{\bibinfo{volume}{5}}, \bibinfo{pages}{328} (\bibinfo{year}{2022}).
\newblock \urlprefix\url{https://doi.org/10.1038/s42005-022-01106-8}.

\bibitem{Xing2024}
\bibinfo{author}{Xing, Y.} \emph{et~al.}
\newblock \bibinfo{title}{Optical manipulation of the charge-density-wave state in rbv3sb5}.
\newblock \emph{\bibinfo{journal}{Nature}} \textbf{\bibinfo{volume}{631}}, \bibinfo{pages}{60--66} (\bibinfo{year}{2024}).
\newblock \urlprefix\url{https://doi.org/10.1038/s41586-024-07519-5}.

\bibitem{Zong2018}
\bibinfo{author}{Zong, A.} \emph{et~al.}
\newblock \bibinfo{title}{{Ultrafast manipulation of mirror domain walls in a charge density wave}}.
\newblock \emph{\bibinfo{journal}{Science Advances}} \textbf{\bibinfo{volume}{4}}, \bibinfo{pages}{eaau5501} (\bibinfo{year}{2018}).

\bibitem{Yoshikawa2021}
\bibinfo{author}{Yoshikawa, N.} \emph{et~al.}
\newblock \bibinfo{title}{{Ultrafast switching to an insulating-like metastable state by amplitudon excitation of a charge density wave}}.
\newblock \emph{\bibinfo{journal}{Nature Physics}} \textbf{\bibinfo{volume}{17}}, \bibinfo{pages}{909--914} (\bibinfo{year}{2021}).

\bibitem{Rettig2021}
\bibinfo{author}{Maklar, J.} \emph{et~al.}
\newblock \bibinfo{title}{{Nonequilibrium charge-density-wave order beyond the thermal limit}}.
\newblock \emph{\bibinfo{journal}{Nature Communications}} \textbf{\bibinfo{volume}{12}}, \bibinfo{pages}{2499} (\bibinfo{year}{2021}).

\bibitem{Vaskivskyi2015}
\bibinfo{author}{Vaskivskyi, I.} \emph{et~al.}
\newblock \bibinfo{title}{{Controlling the metal-to-insulator relaxation of the metastable hidden quantum state in \protect{1T-TaS$_2$}}}.
\newblock \emph{\bibinfo{journal}{Science Advances}} \textbf{\bibinfo{volume}{1}}, \bibinfo{pages}{e1500168} (\bibinfo{year}{2015}).

\bibitem{Forst2011}
\bibinfo{author}{F{\"o}rst, M.} \emph{et~al.}
\newblock \bibinfo{title}{Nonlinear phononics as an ultrafast route to lattice control}.
\newblock \emph{\bibinfo{journal}{Nature Physics}} \textbf{\bibinfo{volume}{7}}, \bibinfo{pages}{854--856} (\bibinfo{year}{2011}).
\newblock \urlprefix\url{https://doi.org/10.1038/nphys2055}.

\bibitem{Meredith2022}
\bibinfo{author}{Henstridge, M.}, \bibinfo{author}{F{\"o}rst, M.}, \bibinfo{author}{Rowe, E.}, \bibinfo{author}{Fechner, M.} \& \bibinfo{author}{Cavalleri, A.}
\newblock \bibinfo{title}{Nonlocal nonlinear phononics}.
\newblock \emph{\bibinfo{journal}{Nature Physics}} \textbf{\bibinfo{volume}{18}}, \bibinfo{pages}{457--461} (\bibinfo{year}{2022}).
\newblock \urlprefix\url{https://doi.org/10.1038/s41567-022-01512-3}.

\bibitem{Galan2020}
\bibinfo{author}{Jiménez-Galán, A.}, \bibinfo{author}{Silva, R. E.~F.}, \bibinfo{author}{Smirnova, O.} \& \bibinfo{author}{Ivanov, M.}
\newblock \bibinfo{title}{{Lightwave control of topological properties in 2D materials for sub-cycle and non-resonant valley manipulation}}.
\newblock \emph{\bibinfo{journal}{Nature Photonics}} \textbf{\bibinfo{volume}{14}}, \bibinfo{pages}{728--732} (\bibinfo{year}{2020}).

\bibitem{Huber2023}
\bibinfo{author}{Ito, S.} \emph{et~al.}
\newblock \bibinfo{title}{{Build-up and dephasing of Floquet--Bloch bands on subcycle timescales}}.
\newblock \emph{\bibinfo{journal}{Nature}} \textbf{\bibinfo{volume}{616}}, \bibinfo{pages}{696--701} (\bibinfo{year}{2023}).

\bibitem{Mitrano2016}
\bibinfo{author}{Mitrano, M.} \emph{et~al.}
\newblock \bibinfo{title}{{Possible light-induced superconductivity in \protect{K$_3$C$_60$} at high temperature}}.
\newblock \emph{\bibinfo{journal}{Nature}} \textbf{\bibinfo{volume}{530}}, \bibinfo{pages}{461--464} (\bibinfo{year}{2016}).

\bibitem{Budden2021}
\bibinfo{author}{Budden, M.} \emph{et~al.}
\newblock \bibinfo{title}{{Evidence for metastable photo-induced superconductivity in \protect{K$_3$C$_60$}}}.
\newblock \emph{\bibinfo{journal}{Nature Physics}} \textbf{\bibinfo{volume}{17}}, \bibinfo{pages}{611--618} (\bibinfo{year}{2021}).

\bibitem{Oka2009}
\bibinfo{author}{Oka, T.} \& \bibinfo{author}{Aoki, H.}
\newblock \bibinfo{title}{Photovoltaic hall effect in graphene}.
\newblock \emph{\bibinfo{journal}{Phys. Rev. B}} \textbf{\bibinfo{volume}{79}}, \bibinfo{pages}{081406} (\bibinfo{year}{2009}).
\newblock \urlprefix\url{https://link.aps.org/doi/10.1103/PhysRevB.79.081406}.

\bibitem{Kitagawa2010}
\bibinfo{author}{Kitagawa, T.}, \bibinfo{author}{Berg, E.}, \bibinfo{author}{Rudner, M.} \& \bibinfo{author}{Demler, E.}
\newblock \bibinfo{title}{{Topological characterization of periodically driven quantum systems}}.
\newblock \emph{\bibinfo{journal}{Physical Review B}} \textbf{\bibinfo{volume}{82}}, \bibinfo{pages}{235114} (\bibinfo{year}{2010}).

\bibitem{Lindner2011}
\bibinfo{author}{Lindner, N.~H.}, \bibinfo{author}{Refael, G.} \& \bibinfo{author}{Galitski, V.}
\newblock \bibinfo{title}{Floquet topological insulator in semiconductor quantum wells}.
\newblock \emph{\bibinfo{journal}{Nature Physics}} \textbf{\bibinfo{volume}{7}}, \bibinfo{pages}{490--495} (\bibinfo{year}{2011}).
\newblock \urlprefix\url{https://doi.org/10.1038/nphys1926}.

\bibitem{Wang2014}
\bibinfo{author}{Wang, R.}, \bibinfo{author}{Wang, B.}, \bibinfo{author}{Shen, R.}, \bibinfo{author}{Sheng, L.} \& \bibinfo{author}{Xing, D.~Y.}
\newblock \bibinfo{title}{Floquet weyl semimetal induced by off-resonant light}.
\newblock \emph{\bibinfo{journal}{Europhysics Letters}} \textbf{\bibinfo{volume}{105}}, \bibinfo{pages}{17004} (\bibinfo{year}{2014}).
\newblock \urlprefix\url{https://dx.doi.org/10.1209/0295-5075/105/17004}.

\bibitem{Gedik2016}
\bibinfo{author}{Mahmood, F.} \emph{et~al.}
\newblock \bibinfo{title}{{Selective scattering between Floquet--Bloch and Volkov states in a topological insulator}}.
\newblock \emph{\bibinfo{journal}{Nature Physics}} \textbf{\bibinfo{volume}{12}}, \bibinfo{pages}{306--310} (\bibinfo{year}{2016}).

\bibitem{Sentef2017}
\bibinfo{author}{Hübener, H.}, \bibinfo{author}{Sentef, M.~A.}, \bibinfo{author}{Giovannini, U.~D.}, \bibinfo{author}{Kemper, A.~F.} \& \bibinfo{author}{Rubio, A.}
\newblock \bibinfo{title}{{Creating stable Floquet–Weyl semimetals by laser-driving of 3D Dirac materials}}.
\newblock \emph{\bibinfo{journal}{Nature Communications}} \textbf{\bibinfo{volume}{8}}, \bibinfo{pages}{13940} (\bibinfo{year}{2017}).

\bibitem{Oka2018}
\bibinfo{author}{Oka, T.} \& \bibinfo{author}{Kitamura, S.}
\newblock \bibinfo{title}{{Floquet Engineering of Quantum Materials}}.
\newblock \emph{\bibinfo{journal}{Annual Review of Condensed Matter Physics}} \textbf{\bibinfo{volume}{10}}, \bibinfo{pages}{1--22} (\bibinfo{year}{2018}).

\bibitem{Rudner2020}
\bibinfo{author}{Rudner, M.~S.} \& \bibinfo{author}{Lindner, N.~H.}
\newblock \bibinfo{title}{{Band structure engineering and non-equilibrium dynamics in Floquet topological insulators}}.
\newblock \emph{\bibinfo{journal}{Nature Reviews Physics}} \textbf{\bibinfo{volume}{2}}, \bibinfo{pages}{229--244} (\bibinfo{year}{2020}).

\bibitem{Oka2016}
\bibinfo{author}{Ebihara, S.}, \bibinfo{author}{Fukushima, K.} \& \bibinfo{author}{Oka, T.}
\newblock \bibinfo{title}{{Chiral pumping effect induced by rotating electric fields}}.
\newblock \emph{\bibinfo{journal}{Physical Review B}} \textbf{\bibinfo{volume}{93}}, \bibinfo{pages}{155107} (\bibinfo{year}{2016}).

\bibitem{Gedik2015}
\bibinfo{author}{Sie, E.~J.} \emph{et~al.}
\newblock \bibinfo{title}{{Valley-selective optical Stark effect in monolayer \protect{WS$_2$}}}.
\newblock \emph{\bibinfo{journal}{Nature Materials}} \textbf{\bibinfo{volume}{14}}, \bibinfo{pages}{290--294} (\bibinfo{year}{2015}).

\bibitem{Matsunaga2023}
\bibinfo{author}{Murotani, Y.} \emph{et~al.}
\newblock \bibinfo{title}{{Disentangling the Competing Mechanisms of Light-Induced Anomalous Hall Conductivity in Three-Dimensional Dirac Semimetal}}.
\newblock \emph{\bibinfo{journal}{Physical Review Letters}} \textbf{\bibinfo{volume}{131}}, \bibinfo{pages}{096901} (\bibinfo{year}{2023}).

\bibitem{Zhou2023}
\bibinfo{author}{Zhou, S.} \emph{et~al.}
\newblock \bibinfo{title}{{Pseudospin-selective Floquet band engineering in black phosphorus}}.
\newblock \emph{\bibinfo{journal}{Nature}} \textbf{\bibinfo{volume}{614}}, \bibinfo{pages}{75--80} (\bibinfo{year}{2023}).

\bibitem{Merboldt2024}
\bibinfo{author}{Merboldt, M.} \emph{et~al.}
\newblock \bibinfo{title}{Observation of \protect{Floquet} states in graphene} (\bibinfo{year}{2024}).
\newblock \urlprefix\url{https://arxiv.org/abs/2404.12791}.
\newblock \eprint{2404.12791}.

\bibitem{Choi2024}
\bibinfo{author}{Choi, D.} \emph{et~al.}
\newblock \bibinfo{title}{Direct observation of \protect{Floquet-Bloch} states in monolayer graphene} (\bibinfo{year}{2024}).
\newblock \urlprefix\url{https://arxiv.org/abs/2404.14392}.
\newblock \eprint{2404.14392}.

\bibitem{Rechtsman2013}
\bibinfo{author}{Rechtsman, M.~C.} \emph{et~al.}
\newblock \bibinfo{title}{Photonic floquet topological insulators}.
\newblock \emph{\bibinfo{journal}{Nature}} \textbf{\bibinfo{volume}{496}}, \bibinfo{pages}{196--200} (\bibinfo{year}{2013}).
\newblock \urlprefix\url{https://doi.org/10.1038/nature12066}.

\bibitem{Gregor2014}
\bibinfo{author}{Jotzu, G.} \emph{et~al.}
\newblock \bibinfo{title}{Experimental realization of the topological haldane model with ultracold fermions}.
\newblock \emph{\bibinfo{journal}{Nature}} \textbf{\bibinfo{volume}{515}}, \bibinfo{pages}{237--240} (\bibinfo{year}{2014}).
\newblock \urlprefix\url{https://doi.org/10.1038/nature13915}.

\bibitem{Monica2024}
\bibinfo{author}{Braun, C.} \emph{et~al.}
\newblock \bibinfo{title}{Real-space detection and manipulation of topological edge modes with ultracold atoms}.
\newblock \emph{\bibinfo{journal}{Nature Physics}} \textbf{\bibinfo{volume}{20}}, \bibinfo{pages}{1306--1312} (\bibinfo{year}{2024}).
\newblock \urlprefix\url{https://doi.org/10.1038/s41567-024-02506-z}.

\bibitem{Zhou2021}
\bibinfo{author}{Bao, C.}, \bibinfo{author}{Tang, P.}, \bibinfo{author}{Sun, D.} \& \bibinfo{author}{Zhou, S.}
\newblock \bibinfo{title}{{Light-induced emergent phenomena in 2D materials and topological materials}}.
\newblock \emph{\bibinfo{journal}{Nature Reviews Physics}} \textbf{\bibinfo{volume}{4}}, \bibinfo{pages}{33--48} (\bibinfo{year}{2022}).

\bibitem{Orth2022}
\bibinfo{author}{Trevisan, T.~V.}, \bibinfo{author}{Arribi, P.~V.}, \bibinfo{author}{Heinonen, O.}, \bibinfo{author}{Slager, R.-J.} \& \bibinfo{author}{Orth, P.~P.}
\newblock \bibinfo{title}{{Bicircular Light Floquet Engineering of Magnetic Symmetry and Topology and Its Application to the Dirac Semimetal \protect{Cd$_3$As$_2$}}}.
\newblock \emph{\bibinfo{journal}{Physical Review Letters}} \textbf{\bibinfo{volume}{128}}, \bibinfo{pages}{066602} (\bibinfo{year}{2022}).

\bibitem{Shimano2024}
\bibinfo{author}{Hirai, Y.}, \bibinfo{author}{Okumura, S.}, \bibinfo{author}{Yoshikawa, N.}, \bibinfo{author}{Oka, T.} \& \bibinfo{author}{Shimano, R.}
\newblock \bibinfo{title}{Floquet weyl states at one-photon resonance: An origin of nonperturbative optical responses in three-dimensional materials}.
\newblock \emph{\bibinfo{journal}{Phys. Rev. Res.}} \textbf{\bibinfo{volume}{6}}, \bibinfo{pages}{L012027} (\bibinfo{year}{2024}).
\newblock \urlprefix\url{https://link.aps.org/doi/10.1103/PhysRevResearch.6.L012027}.

\bibitem{RodriguesVega2019}
\bibinfo{author}{Rodriguez-Vega, M.}, \bibinfo{author}{Kumar, A.} \& \bibinfo{author}{Seradjeh, B.}
\newblock \bibinfo{title}{Higher-order floquet topological phases with corner and bulk bound states}.
\newblock \emph{\bibinfo{journal}{Phys. Rev. B}} \textbf{\bibinfo{volume}{100}}, \bibinfo{pages}{085138} (\bibinfo{year}{2019}).
\newblock \urlprefix\url{https://link.aps.org/doi/10.1103/PhysRevB.100.085138}.

\bibitem{Shimano2022}
\bibinfo{author}{Yoshikawa, N.} \emph{et~al.}
\newblock \bibinfo{title}{{Light-induced chiral gauge field in a massive 3D Dirac electron system}}.
\newblock \emph{\bibinfo{journal}{arXiv}}  (\bibinfo{year}{2022}).

\bibitem{Shimano2023}
\bibinfo{author}{Hirai, Y.} \emph{et~al.}
\newblock \bibinfo{title}{{Anomalous Hall effect of light-driven three-dimensional Dirac electrons in bismuth}}.
\newblock \emph{\bibinfo{journal}{arXiv}}  (\bibinfo{year}{2023}).

\bibitem{Chan2016}
\bibinfo{author}{Chan, C.-K.}, \bibinfo{author}{Lee, P.~A.}, \bibinfo{author}{Burch, K.~S.}, \bibinfo{author}{Han, J.~H.} \& \bibinfo{author}{Ran, Y.}
\newblock \bibinfo{title}{{When Chiral Photons Meet Chiral Fermions: Photoinduced Anomalous Hall Effects in Weyl Semimetals}}.
\newblock \emph{\bibinfo{journal}{Physical Review Letters}} \textbf{\bibinfo{volume}{116}}, \bibinfo{pages}{026805} (\bibinfo{year}{2016}).

\bibitem{vonKlitzing2020}
\bibinfo{author}{von Klitzing, K.} \emph{et~al.}
\newblock \bibinfo{title}{40 years of the quantum \protect{H}all effect}.
\newblock \emph{\bibinfo{journal}{Nature Reviews Physics}} \textbf{\bibinfo{volume}{2}}, \bibinfo{pages}{397--401} (\bibinfo{year}{2020}).
\newblock \urlprefix\url{https://doi.org/10.1038/s42254-020-0209-1}.

\bibitem{Auston1975}
\bibinfo{author}{Auston, D.~H.}
\newblock \bibinfo{title}{{Picosecond optoelectronic switching and gating in silicon}}.
\newblock \emph{\bibinfo{journal}{Applied Physics Letters}} \textbf{\bibinfo{volume}{26}}, \bibinfo{pages}{101--103} (\bibinfo{year}{1975}).

\bibitem{Song2023}
\bibinfo{author}{Ma, Q.}, \bibinfo{author}{Kumar, R.~K.}, \bibinfo{author}{Xu, S.-Y.}, \bibinfo{author}{Koppens, F. H.~L.} \& \bibinfo{author}{Song, J. C.~W.}
\newblock \bibinfo{title}{{Photocurrent as a multiphysics diagnostic of quantum materials}}.
\newblock \emph{\bibinfo{journal}{Nature Reviews Physics}} \textbf{\bibinfo{volume}{5}}, \bibinfo{pages}{170--184} (\bibinfo{year}{2023}).

\bibitem{supplement}
\bibinfo{title}{Supplementary \protect{I}nformation}.

\bibitem{Loh2020}
\bibinfo{author}{Song, P.} \emph{et~al.}
\newblock \bibinfo{title}{{Coexistence of large conventional and planar spin Hall effect with long spin diffusion length in a low-symmetry semimetal at room temperature}}.
\newblock \emph{\bibinfo{journal}{Nature Materials}} \textbf{\bibinfo{volume}{19}}, \bibinfo{pages}{292--298} (\bibinfo{year}{2020}).

\bibitem{Sun2021}
\bibinfo{author}{Le, C.} \& \bibinfo{author}{Sun, Y.}
\newblock \bibinfo{title}{{Topology and symmetry of circular photogalvanic effect in the chiral multifold semimetals: a review}}.
\newblock \emph{\bibinfo{journal}{Journal of Physics: Condensed Matter}} \textbf{\bibinfo{volume}{33}}, \bibinfo{pages}{503003} (\bibinfo{year}{2021}).

\bibitem{Soifer2019}
\bibinfo{author}{Soifer, H.} \emph{et~al.}
\newblock \bibinfo{title}{Band-resolved imaging of photocurrent in a topological insulator}.
\newblock \emph{\bibinfo{journal}{Phys. Rev. Lett.}} \textbf{\bibinfo{volume}{122}}, \bibinfo{pages}{167401} (\bibinfo{year}{2019}).
\newblock \urlprefix\url{https://link.aps.org/doi/10.1103/PhysRevLett.122.167401}.

\bibitem{Chan2017a}
\bibinfo{author}{Chan, C.-K.}, \bibinfo{author}{Lindner, N.~H.}, \bibinfo{author}{Refael, G.} \& \bibinfo{author}{Lee, P.~A.}
\newblock \bibinfo{title}{Photocurrents in weyl semimetals}.
\newblock \emph{\bibinfo{journal}{Phys. Rev. B}} \textbf{\bibinfo{volume}{95}}, \bibinfo{pages}{041104} (\bibinfo{year}{2017}).
\newblock \urlprefix\url{https://link.aps.org/doi/10.1103/PhysRevB.95.041104}.

\bibitem{Golub2022}
\bibinfo{author}{Leppenen, N.~V.} \& \bibinfo{author}{Golub, L.~E.}
\newblock \bibinfo{title}{{Nonlinear optical absorption and photocurrents in topological insulators}}.
\newblock \emph{\bibinfo{journal}{Physical Review B}} \textbf{\bibinfo{volume}{105}}, \bibinfo{pages}{115306} (\bibinfo{year}{2022}).

\bibitem{SuYang2018}
\bibinfo{author}{Xu, S.-Y.} \emph{et~al.}
\newblock \bibinfo{title}{{Electrically switchable Berry curvature dipole in the monolayer topological insulator \protect{WTe$_2$}}}.
\newblock \emph{\bibinfo{journal}{Nature Physics}} \textbf{\bibinfo{volume}{14}}, \bibinfo{pages}{900--906} (\bibinfo{year}{2018}).

\bibitem{Liuyan2021}
\bibinfo{author}{Tiwari, A.} \emph{et~al.}
\newblock \bibinfo{title}{{Giant c-axis nonlinear anomalous Hall effect in \protect{T$_d$-MoTe$_2$ and WTe$_2$}}}.
\newblock \emph{\bibinfo{journal}{Nature Communications}} \textbf{\bibinfo{volume}{12}}, \bibinfo{pages}{2049} (\bibinfo{year}{2021}).

\bibitem{Sodeman2021}
\bibinfo{author}{Matsyshyn, O.}, \bibinfo{author}{Piazza, F.}, \bibinfo{author}{Moessner, R.} \& \bibinfo{author}{Sodemann, I.}
\newblock \bibinfo{title}{{Rabi Regime of Current Rectification in Solids}}.
\newblock \emph{\bibinfo{journal}{Physical Review Letters}} \textbf{\bibinfo{volume}{127}}, \bibinfo{pages}{126604} (\bibinfo{year}{2021}).

\bibitem{Nagaosa2016}
\bibinfo{author}{Morimoto, T.} \& \bibinfo{author}{Nagaosa, N.}
\newblock \bibinfo{title}{{Topological nature of nonlinear optical effects in solids}}.
\newblock \emph{\bibinfo{journal}{Science Advances}} \textbf{\bibinfo{volume}{2}}, \bibinfo{pages}{e1501524} (\bibinfo{year}{2016}).

\bibitem{Bakos1977}
\bibinfo{author}{Bakos, J.}
\newblock \bibinfo{title}{{AC stark effect and multiphoton processes in atoms}}.
\newblock \emph{\bibinfo{journal}{Physics Reports}} \textbf{\bibinfo{volume}{31}}, \bibinfo{pages}{209--235} (\bibinfo{year}{1977}).

\bibitem{Grioni2017}
\bibinfo{author}{Crepaldi, A.} \emph{et~al.}
\newblock \bibinfo{title}{{Enhanced ultrafast relaxation rate in the Weyl semimetal phase of \protect{MoTe$_2$} measured by time- and angle-resolved photoelectron spectroscopy}}.
\newblock \emph{\bibinfo{journal}{Physical Review B}} \textbf{\bibinfo{volume}{96}}, \bibinfo{pages}{241408} (\bibinfo{year}{2017}).

\bibitem{BiswasSoren2021}
\bibinfo{author}{Biswas, D.} \emph{et~al.}
\newblock \bibinfo{title}{{Ultrafast Triggering of Insulator--Metal Transition in Two-Dimensional \protect{VSe$_2$}}}.
\newblock \emph{\bibinfo{journal}{Nano Letters}} \textbf{\bibinfo{volume}{21}}, \bibinfo{pages}{1968--1975} (\bibinfo{year}{2021}).

\bibitem{Moore2017}
\bibinfo{author}{de~Juan, F.}, \bibinfo{author}{Grushin, A.~G.}, \bibinfo{author}{Morimoto, T.} \& \bibinfo{author}{Moore, J.~E.}
\newblock \bibinfo{title}{Quantized circular photogalvanic effect in weyl semimetals}.
\newblock \emph{\bibinfo{journal}{Nature Communications}} \textbf{\bibinfo{volume}{8}}, \bibinfo{pages}{15995} (\bibinfo{year}{2017}).
\newblock \urlprefix\url{https://doi.org/10.1038/ncomms15995}.

\bibitem{pizzi2020}
\bibinfo{author}{Pizzi, G.} \emph{et~al.}
\newblock \bibinfo{title}{{Wannier90 as a community code: new features and applications}}.
\newblock \emph{\bibinfo{journal}{J. Condens. Matter Phys.}} \textbf{\bibinfo{volume}{32}}, \bibinfo{pages}{165902} (\bibinfo{year}{2020}).

\bibitem{Sato2019}
\bibinfo{author}{Sato, S.~A.} \emph{et~al.}
\newblock \bibinfo{title}{{Microscopic theory for the light-induced anomalous Hall effect in graphene}}.
\newblock \emph{\bibinfo{journal}{Physical Review B}} \textbf{\bibinfo{volume}{99}}, \bibinfo{pages}{214302} (\bibinfo{year}{2019}).

\bibitem{Mathey2020}
\bibinfo{author}{Nuske, M.} \emph{et~al.}
\newblock \bibinfo{title}{{Floquet dynamics in light-driven solids}}.
\newblock \emph{\bibinfo{journal}{Physical Review Research}} \textbf{\bibinfo{volume}{2}}, \bibinfo{pages}{043408} (\bibinfo{year}{2020}).

\bibitem{Merlin2023}
\bibinfo{author}{Merlin, R.}
\newblock \bibinfo{title}{Unraveling the effect of circularly polarized light on reciprocal media: Breaking time reversal symmetry with non-maxwellian magnetic-esque fields} (\bibinfo{year}{2023}).
\newblock \urlprefix\url{https://arxiv.org/abs/2309.13622}.
\newblock \eprint{2309.13622}.

\end{thebibliography}

\end{document}